\documentclass{article}

\usepackage{amssymb,amsfonts,amsmath,stmaryrd}
\usepackage{cite,enumerate,float,indentfirst}
\usepackage{color}
\usepackage{tikz}
\usetikzlibrary{arrows,snakes,backgrounds}

\def\be{\begin{eqnarray}}
\def\ee{\end{eqnarray}}
\def\nn{\nonumber}

\def\tr{{\rm tr}\,}
\def\Tr{{\rm Tr}\,}

\definecolor{red}{rgb}{1,0,0}
\definecolor{orange}{rgb}{1,0.5,0}
\definecolor{violet}{rgb}{0.7,0,1}

\hoffset= - 1.0in         

\begin{document}

\begin{center}
\begin{small}
\hfill MIPT/TH-04/22\\
\hfill FIAN/TD-03/22\\
\hfill ITEP/TH-06/22\\
\hfill IITP/TH-05/22\\

\end{small}
\end{center}

\vspace{.5cm}

\begin{center}
\begin{Large}\fontfamily{cmss}
\fontsize{15pt}{27pt}
\selectfont
	\textbf{New Insights into Superintegrability from Unitary Matrix  Models}
	\end{Large}
	
\bigskip \bigskip

\begin{large}
A. Mironov$^{a,b,c,}$\footnote{mironov@lpi.ru; mironov@itep.ru},
A. Morozov$^{d,b,c,}$\footnote{morozov@itep.ru}\ \ and
Z. Zakirova$^{c,e,}$\footnote{zolya\_zakirova@mail.ru}
\end{large}

\bigskip

\begin{small}
$^a$ {\it Lebedev Physics Institute, Moscow 119991, Russia}\\
$^b$ {\it Institute for Information Transmission Problems, Moscow 127994, Russia}\\
$^c$ {\it NRC ``Kurchatov Institute" - ITEP, Moscow 117218, Russia}\\
$^d$ {\it MIPT, Dolgoprudny, 141701, Russia}\\
$^e$ {\it Kazan State Power Engineering University, Kazan, Russia}
\end{small}
 \end{center}

\bigskip

\begin{abstract}
Some eigenvalue matrix models possess an interesting property:
one can manifestly define the basis where all averages can be explicitly calculated.
For example, in the Gaussian Hermitian and rectangular complex models, averages of the
Schur functions are again expressed through the Schur functions.
However, so far this property remains restricted to very particular (e.g. Gaussian) measures.
In this paper, we extend this observation to unitary matrix integrals,
where one could expect that this restriction is easier to lift.
We demonstrate that this is indeed the case,
only this time the Schur averages are linear combinations of the Schur functions.
Full factorization to a single item in the sum
appears only on the Miwa locus, where at least one half of the time-variables
is expressed through matrices of the same size.
For unitary integrals, this is a manifestation of the de Wit-t'Hooft anomaly,
which prevents the answer to be fully analytic in the matrix size $N$.
Once achieved, this understanding can be extended back to the
Hermitian model, where the phenomenon looks very similar:
beyond Gaussian measures superintegrability requires an additional summation.
\end{abstract}

\bigskip

\bigskip

\section{Introduction}

Matrix models play a special role in theoretical physics,
because they are the simplest prototypes of the full string theory.
As any integrals, their partition functions are invariant under the
change of integration variables, but, in this case, the corresponding
Ward identities are especially simple, and acquire the form of
linear Virasoro-like constraints, which further imply a hidden integrability,
i.e. bilinear Hirota-type identities \cite{UFN3}.
In fact, this is not the end of the story: the most interesting classes
of matrix models possess an even stronger feature of {\it superintegrability} \cite{MMsum},
that is, in a certain full basis all the correlation functions are explicitly
calculable (see many examples in \cite{DiF}--\cite{MMMZh}, and also some preliminary results in \cite{Kaz}--\cite{MKR}).
This resembles emergence of closed orbits for the motion in the Coulomb
and harmonic oscillator fields, and the name {\it superintegrability} is borrowed
from those well-known examples.
So far, this property was found for peculiar classes of matrix models,
see \cite{MMrev} for a brief review.
The unitary matrix models did not belong to the list in that summary,
and in this paper we explain what is the difference.
In unitary models, additional time-variables $\bar p_k$,
which parameterize the background potential, appear on equal footing with
the ordinary couplings $p_k$, which are used to describe arbitrary correlators.
This gives a chance to go beyond Gaussian models, towards arbitrary
Dijkgraaf-Vafa phases and even further.

In this paper, we study this possibility and demonstrate that,
in general, the averages are not just factorized, but are represented
as series, which get reduced to single terms only at the Miwa locus.
Moreover, the formulas are not fully analytic in the matrix size $N$,
a well known phenomenon for unitary integrals,
sometime called the de Wit-t'Hooft anomaly.
It also appears straightforward to describe not only the character phase,
to which our formalism is {\it a priori} tuned,
but also the ``opposite" Kontsevich phase,
i.e. a kind of a non-perturbative expansion.

After this peculiarity of superintegrability is understood,
we can return to the Hermitian matrix model and obtain the sum,
which generalizes the usual factorized expressions of Gaussian averages.
A way to cleverly handle such sums and to restore analyticity in $N$
remains for future analysis.

\paragraph{Notation.}

In this letter, one of the main objects are the Schur functions $S_R(x_i)$ \cite{Mac}, which are symmetric functions of variables $x_i$, and are labelled by partitions $R$. One can equivalently consider them as graded polynomials of power sums $p_k:=\sum_i x_i^k$, we use the notation $S_R\{p_k\}$ in this case. For an $N\times N$ matrix $X$ with the eigenvalues $x_i$ so that $p_k=\Tr X^k$ we will also use the notation $S_R[X]$. For the partition $R$, we denote through $l_R$ the number of its parts. Definitions of other functions used in the letter can be found in the short Appendix.

\section{The unitary matrix model}

The main quantity of our interest in this paper is the partition function of unitary matrix model \cite{UMM,KMMOZ}
\be
Z_N\{p,\bar p\}:=\int [DU] \exp\left(\sum_k{p_k\Tr U^k\over k}+{\bar p_k\Tr\! \left(U^\dag\right)^k\over k}\right)
\ee
where $U$ is $N\times N$ unitary matrix and $[dU]$ denotes the Haar measure on unitary matrices. Since the integrand depends only on invariant quantities, one can integrate over angular variables to get an eigenvalue model reducing the partition function to an integral over the eigenvalues $z_i$ of $U$, which are unimodular numbers, $\bar z_i = \frac{1}{z_i}$. We normalize the measure $DU$ in a way that it includes the normalization factor (the volume of the unitary group $U(N)$) emerging from this integration over the angular variables.
Then, the eigenvalue representation of the partition function gets the form \cite{Mehta}
\be
Z_N\{p,\bar p\}=\prod_{i=1}^N\oint_{|z_i|=1}{dz_i\over z_i}\Big|\Delta(z)\Big|^2
\exp\left(\sum_k{p_kz_i^k\over k}+{\bar p_k\over kz_i^k}\right)
=\det_{1\le i,j\le N}C_{i-j}\{p,\bar p\}
\label{detZ}
\ee
where $\Delta(z):=\prod_{i<j}(z_i-z_j)$ is the Vandermonde determinant, and the moment matrix
\be\label{UC}
C_k\{p,\bar p\}:=\oint {dz\over z}z^k\exp\left(\sum_k{p_kz^k\over k}+{\bar p_k\over kz^k}\right)
\ee
Making use of the formula
\be\label{h}
\exp\Big(\sum_k {p_kz^k\over k}\Big)=\sum_m{S_{[m]}\{p\}z^m}
\ee
we can shift the $p,\bar p$-dependence to symmetric Schur polynomials $S_{[m]}$:
\be
Z_N\{p,\bar p\}=\det_{1\le i,j\le N}C_{i-j}\{p,\bar p\}=
\det_{1\le i,j\le N}\sum_{k,l} S_{[k]}\{p\}S_{[l]}\{\bar p\}C_{i-j+k-l}\{0,0\} =
\nn
\ee
\be
=\det_{1\le i,j\le N}\sum_{k,l} S_{[j+k]}\{ p\}S_{[i+l]}\{\bar p\}C_{k-l}\{0,0\}
\ee

\section{Superintegrability of the unitary matrix model}

In fact, since $C_{k}\{0,0\}=\delta_{k,0}$, one can further rewrite this sum as
(see \cite[Eq.(7)]{MMMR})
\be
\det_{1\le i,j\le N}\sum_{k} S_{[j+k]}\{p\}S_{[i+k]}\{\bar p\}
\stackrel{CB}{=}\!\!\!\!\!\! \sum_{k_1>k_2>\ldots>k_N}
\det_{1\le i,j\le N}S_{[k_i+j]}\{p\}\cdot\det_{1\le i,j\le N}S_{[k_i+j]}\{\bar p\}
= \!\! \sum_{R:\ l_R\le N}S_R\{p\}S_R\{\bar p\}
\ee
where we used  the Cauchy-Binet formula (CB),
\be
\det_{1\le i,j\le N} \left(\sum_k A_{ik}B_{kj}\right) \ \stackrel{CB}{=}
\sum_{k_1>k_2>\ldots>k_N}
\det_{1\le i,j\le N}A_{ik_j} \cdot \det_{1\le i,j\le N}B_{ik_j}
\ee
In the last transition in (\ref{main}) we changed the variables $k_i:=R_i-i$,
what allows us to represent the  sum as going over partitions $R$.
We also used the first Jacobi-Trudi identity
\be\label{JT}
S_R\{p\}=\det_{i,j}S_{[R_i-i+j]}\{p\}
\ee
to substitute determinants of symmetric Schur polynomials as generic Schur functions. Thus, we finally obtain
\be\label{main}
Z_N\{p,\bar p\}=\int [DU] \exp\left(\sum_k{p_k\Tr U^k\over k}+{\bar p_k\Tr\! \left(U^\dag\right)^k\over k}\right)=\sum_{R:\ l_R\le N}S_R\{p\}S_R\{\bar p\}
\ee

\bigskip

Eq.(\ref{main}) implies that
\be
\left<\exp\left(\sum_k P_k\Tr U^k\right)\right>=\sum_{R:\ l_R\le N}S_R\{p+P\}S_R\{\bar p\}=
\sum_QS_Q\{P\}\sum_{R:\ l_R\le N}S_{R/Q}\{p\}S_R\{\bar p\}
\ee
and since
\be
\left<\exp\left(\sum_k P_k\Tr U^k\right)\right>=\sum_QS_Q\{P\}\left<S_Q[U]\right>
\ee
this allows us to get an average of the Schur function:
\be\label{scor}
\Big<S_Q[U]\Big>=\sum_{R:\ l_R\le N}S_{R/Q}\{p\}S_R\{\bar p\}
\ee
Likewise,
\be
\boxed{
\Big<S_{Q_1}[U]S_{Q_2}[U^\dag]\Big>=\sum_{R:\ l_R\le N}S_{R/Q_1}\{p\}S_{R/Q_2}\{\bar p\}
}
\label{avetwo}
\ee
This sum goes over infinite set of arbitrary large partitions $R$,
still it is always convergent.
This is not that surprising if one recalls the Cauchy identity for the skew Schur functions
\be\label{C}
\sum_R S_{R/Q_1}\{p\}S_{R/Q_2}\{\bar p\}=\exp\left(\sum_k{p_k\bar p_k\over k}\right)
\sum_\sigma S_{Q_1/\sigma}\{\bar p\}S_{Q_2/\sigma}\{p\}
\ee
This identity expresses the infinite sum over $R$ at the l.h.s. via the finite sum over $\sigma$ at the r.h.s.,
and in (\ref{avetwo}) this infinite sum is further restricted to $l_R\leq N$.

\section{Reduction to special Miwa locus\label{Miwas}}

Let us note that the restriction of sums over partitions, $l_R\le N$ like that in formula (\ref{main}) can be automatically fulfilled of one performs a Miwa transform of the variables $p_k$ and restricts the number of Miwa variables by $N$:
\be\label{Miwa}
\bar p_k=\sum_{i=1}^Nx_i^k
\ee
This is what happens in the Hermitian matrix model: a similar sum for its partition function \cite{MM} is not specifically restricted, since the summand contains $S_R\{\bar p_k=N\}$ instead of $S_R\{\bar p_k\}$ with arbitrary $\bar p_k$. The condition $\bar p_k=N$ means there are only $N$ Miwa variables, $x_i=1$, $i\le N$, and this condition emerges automatically in the model. On contrary, such a condition is not obligatory imposed in the unitary model case, which gives a more generic example, but at the price of explicit restriction on the summation domain. However, if one restricts $\bar p_k$ to only $N$ (arbitrary) Miwa variables (\ref{Miwa}), it gives
\be
Z_N\{p,x\}=\sum_{R}S_R\{p\}S_R(x_i)=\exp\left(\sum_{k,i}{p_kx_i^k\over k}\right)
\ee
One can also realize this Miwa transform by an $N\times N$ matrix $X$ such that $\bar p_k=\Tr X^k$:
\be
Z_N\{p,X\}=\int [DU] \exp\left(\sum_k{p_k\Tr U^k\over k}+{\Tr X^k\Tr\! \left(U^\dag\right)^k\over k}\right)=
\sum_{R}S_R\{p\}S_R[X]=\exp\left(\sum_{k}{p_k\Tr X^k\over k}\right)
\ee
and similarly for (\ref{avetwo}). In particular, one can generate from this formula arbitrary correlators of positive powers of $U$ in this restricted background of $U^{-1}$ (it can be also obtained from (\ref{scor}) using (\ref{C})):
\be
 \Big<S_{Q}[U]\Big>=S_Q[X]\exp\left(\sum_{k}{p_k\Tr X^k\over k}\right)
\ee
or
\be\label{fact}
\boxed{
\Big<\hspace{-.13cm}\Big< S_{Q}[U]\Big>\hspace{-.13cm}\Big>=S_Q[X]
}
\ee
for the normalized correlators. This brings us back to the factorized form of the Schur averages without summation like the Hermitian and complex model cases \cite{MM}.

In fact, one can naturally consider even a more restrictive case of $p_k$ also parameterized by $N$ Miwa variables: $p_k=\sum_iy_i^k$ (or $p_k=\Tr Y^k$), it gives
\be
Z_N\{y,x\}=\sum_{R}S_R(y_i)S_R(x_i)=\prod_{i,j}^N{1\over 1-x_iy_j}
\ee
Thus, the restriction to $N$ Miwa variables makes the expressions too trivial.

\section{A more general model: switching on background fields}

Consider now a more general model with the partition function depending on two external $N\times N$ matrices $A$ and $B$:
\be\label{gBGW}
Z_N\{A,B,p,\bar p\}:=\int [DU] \exp\left(\sum_k{p_k\Tr (AU)^k\over k}+{\bar p_k\Tr \left(U^\dag B\right)^k\over k}\right)
\ee
In fact, this model depends on the matrix of product $AB$ only,
this follows from the change of variables $U\to BU$ and invariance
of the Haar measure under this transformation.

In order to solve this model, we use the Cauchy formula in order to expand the r.h.s. into the sum of $S_R\{p\}S_Q\{\bar p\}$ and integrate the coefficients of expansion over $U$ using the formula \cite{Mac}
\be
\int [DU] S_{R}[AU]S_{Q}[U^\dag B]={S_R[AB]\over S_R\{N\}}\delta_{R,Q}
\label{intSS}
\ee
In fact, this formula can be used as an alternative to the calculation in (\ref{main}).
Its application to (\ref{gBGW}) provides the following result:
\be
Z_N\{A,B,p,\bar p\}=\sum_{P,Q}S_R\{p\}S_Q\{\bar p\}{S_R[AB]\over S_R\{N\}}\delta_{R,Q}=
\sum_{R:\ l_R\le N}S_R\{p\}S_R\{\bar p\}{S_R[AB]\over S_R\{N\}}
\ee
which reduces to (\ref{main}) at $A=B=I$. Similarly to our calculation in the previous subsection, one can immediately obtain now the corresponding correlation functions:
\be
\boxed{
\Big<S_{Q_1}[AU]S_{Q_2}[U^\dag B]\Big>=\sum_{R:\ l_R\le N}S_{R/Q_1}\{p\}S_{R/Q_2}\{\bar p\}{S_R[AB]\over S_R\{N\}}
}
\ee
This expression looks like having poles at integer values of $N$, since
\be\label{dim}
S_R\{N\}=\prod_{i,j\in R}(N-i+j)\cdot S_R\{p_k=\delta_{k,1}\}
\ee
However, it turns out that these poles do not emerge due to the restriction of the sum to the partitions of lengths not larger than $N$. Indeed, as follows from (\ref{dim}), the pole in $N$ that may come from the diagram $R$ is at most at $l_R-1$. Hence, the restriction means that contributes only the diagrams that do not give rise to a pole at a given $N$. This simultaneously means that there is no analytic expression in $N$  possible for the correlators in this model. This is called de Wit -- t'Hooft anomaly \cite{DH}.

\section{BGW model}

Choosing $p_k=\delta_{k,1}$, $\bar p_k=\delta_{k,1}$ gives us the Brez\'in-Gross-Witten (BGW) model \cite{BGW} in the character phase, i.e. perturbatively in $AB$ \cite{GKMU}. The BGW partition function looks like \cite{Bars,GKMU}
\be\label{BGW-}
Z_N^{BGW}\{A \}:=
\int [DU] e^{\Tr A(U + U^{-1})} =
\sum_{R:\ l_R\le N}{S_R\{\delta_{k,1}\}^2\over S_R\{N\}}S_R[A^2]
\ee
and the correlators are
 \be
 \boxed{
\Big<S_{Q_1}[AU]S_{Q_2}[AU^\dag]\Big>=
\sum_{R:\ l_R\le N}S_{R/Q_1}\{\delta_{k,1}\}S_{R/Q_2}\{\delta_{k,1}\}{S_R[A^2]\over S_R\{N\}}
}
\ee
Using the formula that expresses the skew Schur functions via the shifted Schur functions $S^*_Q$ at the locus $p_k=\delta_{k,1}$, \cite{Ok},
\be
S_{R/Q}\{\delta_{k,1}\}=S_R\{\delta_{k,1}\}\cdot S^*_Q(R_i)
\ee
one can also rewrite this average in the form
\be
\Big<S_{Q_1}[AU]S_{Q_2}[AU^\dag]\Big>=\sum_{R:\ l_R\le N}S^*_{Q_1}(R_i)S^*_{Q_2}(R_i)
{S_R\{\delta_{k,1}\}^2\over S_R\{N\}}S_R[A^2]
\ee
The quantity $S_R\{\delta_{k,1}\}$ is often denoted by $d_R$.

\section{BGW model in Kontsevich phase}

An interesting question is how to get an answer in the Kontsevich case \cite{GKMU}, i.e. the integral
\be
Z_N^{BGW}\{A \}:=
\int [DU] e^{\Tr A(U + U^{-1})}
\ee
at large $A$. For the sake of simplicity, from now on, we assume $A$ is a Hermitian matrix with positive trace. In order for this integral expansion at large $A$ to start from 1, we use a peculiar normalization of the integral \cite{GKMU}
\be
Z_N^{BGW}\{A \}_+:=e^{-2\Tr A}\sqrt{\det (A\otimes A^2+A^2\otimes A)\over (\det A)^N}
\int [DU] e^{\Tr A(U + U^{-1})}
\ee
Hereafter, we label the partition function $Z_N^{BGW}\{A \}$ at large values of $A$, i.e. in the Kontsevich phase, by the subscript $+$, and, at small values, i.e. in the character phase and without the peculiar normalization factor, by the subscript $-$.

\subsection{The case of $N=1$}

It is instructive to begin from the $N=1$ example, where the BGW partition function reduces
to a (modified) Bessel function with well known expansions in both positive (character phase)
and negative (Kontsevich phase) powers of  $\lambda$.

Indeed, in this case,
\be
Z_{N=1}^{BGW}\{A \}_-=\int_0^{2\pi} {d\phi\over 2\pi} e^{2A\cos\phi}=I_0(2A)
\ee
and the series in $A$ for the modified Bessel function \cite{GR} gives
\be
Z_{N=1}^{BGW}\{A \}_-=\sum_{k=0}{A^{2k}\over k!^2}
\ee
which coincides with (\ref{BGW-}) at $N=1$.

Similarly, using the asymptotic expansion of the modified Bessel function at large $A$ \cite{GR}, one obtains
\be
Z_{N=1}^{BGW}\{A \}_+=2\sqrt{A\pi}e^{-2A}I_0(2A)=\sum_{k=0} {(2k)!^2\over k!^3}\Big({1\over 64A}\Big)^k=
\sum_{k=0}{(-1)^k\over k!}{\Gamma(1/2+k)\over\Gamma(1/2-k)}\Big({1\over 4A}\Big)^k
\ee
However, the main point is not yet seen at the level of $N=1$.

\subsection{The case of $N=2$:  $Q$ Schur versus Schur functions}

The most interesting aspect of the story is that expansions at the two phases
are drastically different:
one, in the character phase, is in Schur functions,
while another one, in the Kontsevich phase, is reduced to a smaller set
of $Q$-Schur functions \cite{Schur,Mac}.
To see this difference, we should proceed to $N=2$.
Then, in the character phase, the expansion of the BGW partition function is
\be
Z_{N=2}^{BGW}\{A \}_-=1+{1\over 2}(a_1^2+a_2^2)+{1\over 12}(a_1^4+4a_1^2a_2^2+a_2^4)=1+{1\over 2}p_1
+{1\over 6}p_1^2-{1\over 12}p_2
\ee
depends on {\it all} $p_k = a_1^{2k}+a_2^{2k}$ and can be expanded into the basis of the Schur functions.
At the same time, in the Kontsevich phase (see (\ref{GKMdet})-(\ref{Bes}) below),
\be
Z_{N=2}^{BGW}\{A \}_+={\psi_1(a_1)\psi_2(a_2)-\psi_1(a_2)\psi_2(a_1)\over a_1-a_2}
=1+{1\over 16}\left({1\over a_1}+
{1\over a_2}\right)+{9\over 512}\left({1\over a_1^2}+{2\over a_1a_2}+{1\over a_2^2}\right)+\ldots
=\nn\\=1+{1\over 16}\tilde p_1+{9\over 512}\tilde p_1^2+\ldots
\ee
the dependence on $\tilde p_2: = a_1^{-2} + a_2^{-2}$ disappears.
Looking at higher terms in the expansion, one observes that actually
{\it all} the variables with even indices, $\tilde p_{2k}$ drop out.
This implies that, in the Kontsevich phase, one has to use a restricted basis of symmetric polynomials that depend only on $p_k$ with odd $k$. Such a basis is known, these are Q Schur functions \cite{Schur,Mac}.

\subsection{Generic integer $N$}

The simplest way to obtain the expansion of $Z_N^{BGW}\{A \}_+$ at generic $N$ is to use a determinant representation \cite{GKMU}:
\be\label{GKMdet}
Z_N^{BGW}\{A \}_+={\det_{i,j}\psi_i(a_j)\over\Delta(a)}
\ee
where
\be
\psi_i(a):=2\sqrt{\pi}a^{i-1/2}e^{-2a}I_{i-1}(2a)
\ee
and we denoted through $a_i$ the eigenvalues of the matrix $A$. As we already pointed out, an important property of (\ref{GKMdet}) is that it is a function of traces of only odd degrees of the matrix $A$: $\tilde p_{2k-1}=\Tr A^{-2k+1}$. Now, using the asymptotics of the modified Bessel function \cite{GR}
\be\label{Bes}
I_n(2a)={e^{2a}\over 2\sqrt{\pi a}}\sum_{k=0}{(-1)^k\over k!}{\Gamma(n+1/2+k)\over\Gamma(n+1/2-k)}\Big({1\over 4A}\Big)^k
\ee
one obtains that
\be
\boxed{
Z_N^{BGW}\{A \}_+={\det_{i,j}\psi_i(a_j)\over\Delta(a)}=
\sum_{R\in SP}\left({1\over 32}\right)^{|R|}Q_{R}\{\Tr A^{-k}\}{Q_{R}\{\delta_{k,1}\}^3\over Q_{2R}\{\delta_{k,1}\}^2}
}
\ee
where $Q_R$ are the Q Schur functions, and $SP$ means strict partitions, i.e. those with all lines of distinct lengths. This formula was conjectured in \cite{Alex} and later proved in \cite{ACh}.

\section{Generalized Itzykson-Zuber model}

One can also consider a generalized Itzykson-Zuber integral:
\be
Z^{IZ}_N\{A,B,p\}:=\int [DU] \exp\left(\sum_k{p_k\Tr (AUBU^\dag)^k\over k}\right)
\label{IZgen}
\ee
This model can be also solved in the same way using the formula for the integral \cite{Mac}
\be
\int [DU] S_R[AUBU^\dag] ={S_R[A]S_R[B]\over S_R\{N\}}
\ee
which is a kind of a ``dual" to (\ref{intSS}).
Applying the Cauchy identity (\ref{C}) to (\ref{IZgen}), we obtain
\be
Z^{IZ}_N\{A,B,p\}=\sum_RS_R\{p\}\int [DU] S_R[AUBU^\dag]=\sum_{R:\ l_R\le N}S_R\{p\}{S_R[A]S_R[B]\over S_R\{N\}}
\ee
Using the same trick as before, one can obtain  the correlators in the form
\be
\boxed{
\Big<S_{Q}[AUBU^\dag]\Big>^{IZ}:=
\int [DU] S_{Q}[AUBU^\dag] \exp\left(\sum_k{p_k\Tr (AUBU^\dag)^k\over k}\right)
=\sum_{R:\ l_R\le N}S_{R/Q}\{p\}{S_R[A]S_R[B]\over S_R\{N\}}
}
\ee
In fact, this sum is assumed to be automatically reduced to $l_R\leq N$,
since $N$ is also the size of the background matrices $A$ and $B$,
and two zeroes in the numerator overweight a single zero in denominator. Hence, the restriction on summation can be omitted.

\section{Back to the case of Hermitian model}

In the case of Hermitian model \cite{Mehta,KMMOZ},
\be
{\cal Z}_N\{p\}:=\int \rho(H) dH \exp\left(\sum_k{p_k\Tr H^k\over k} \right)
=\prod_{i=1}^N\int \rho(h_i){dh_i}\cdot \Delta^2(h)
\exp\left(\sum_k{p_kh_i^k\over k} \right)
=\det_{1\le i,j\le N}{\cal C}_{i+j-2}\{p \}
\label{detZH}
\ee
where $\rho(h)$ is a measure, and
\be\label{UCH}
{\cal C}_k\{p \}:=\int h^k\rho(h) dh \exp\left(\sum_k{p_kh^k\over k} \right)
\ee
one  could make just the same calculation as in (\ref{main}),
see sec.2.2 of \cite{MMMR}:
\be\label{mainH}
Z_N\{p \}=\det_{1\le i,j\le N}{\cal C}_{i+j-2}\{p,\bar p\}=(-1)^{N(N-1)\over 2}\det_{1\le i,j\le N}{\cal C}_{N-i+j-1}\{p_k\}=\nn\\
=(-1)^{N(N-1)\over 2}\det_{1\le i,j\le N}\sum_k S_{[k]}\{p_k\}{\cal C}_{N-i+j+k-1}\{0\}=\nn\\
=(-1)^{N(N-1)\over 2}\det_{1\le i,j\le N}\sum_k S_{[k+i]}\{p_k\}{\cal C}_{N+j+k-1}\{0\}\stackrel{CB}{=}\nn\\
\stackrel{CB}{=}(-1)^{N(N-1)\over 2}\sum_{k_1>k_2>\ldots>k_N}
\det_{1\le i,j\le N}S_{[k_i+j]}\{p_k\}\cdot\det_{1\le i,j\le N}{\cal C}_{N+j+k_i-1}\{0\}=\nn\\
=(-1)^{N(N-1)\over 2}\sum_{R:\ l_R\le N}\det_{1\le i,j\le N}\underbrace{{\cal C}_{N-i+j+R_i-1}}_{{\cal C}_R}\{0\}\cdot S_R\{p_k\}
\ee
The problem is that, in this case, there are no negative degrees of the matrix in the potential in (\ref{detZH}) (otherwise, the matrix integral is not well-defined), and ${\cal C}_R$ is now a sophisticated function of $R$:
\be
{\cal C}_R = \int x^{N-i+j+R_i-1} \rho(h)dh
\ee
Moreover, the factor $\rho(h)$ can not be ignored: one can not put $\rho(h)=1$, as in (\ref{main},
because then the integral over the eigenvalue $h$ diverges.
In fact, ${\cal C}_R$ can be explicitly calculated for the Gaussian measure $\rho(h)=e^{-h^2/2}$
and is expressed through the Schur functions \cite{MM}.
Moreover, as we already mentioned in sec.\ref{Miwas},
this Gaussian ${\cal C}_R$ explicitly vanishes for $l_R>N$, hence the sum in (\ref{mainH})
can be extended to all $R$, so that particular averages get factorized.
This is how the superintegrability phenomenon looks in the Hermitian case.
The situation is similar in the complex rectangular model with Gaussian measure \cite{MM}.

In the unitary case, one does not need $\rho(h)\neq 1$, but the answer involves a non-trivial restriction $l_R\leq N$
in the sum beyond the restricted Miwa locus of sec.\ref{Miwas}.
Still this extension is now quite explicit and easily handleable,
not like (\ref{mainH}) for non-Gaussian $\rho$.

Note that we have come closer to the situation with the Hermitian model by restricting $\bar p_k$ to a peculiar Miwa locus (\ref{Miwa}) in sec.\ref{Miwas} above: then unrestricted $p_k$ can be used to generate arbitrary correlators of positive powers of $U$ in a
restricted background of $U^{-1}$, (\ref{fact}).
Not surprisingly, further restriction of $p_k$ to a similar peculiar Miwa locus makes the partition function elementary and is not interesting. An advantage of the unitary model is a possibility of releasing both $p_k$ and $\bar p_k$ from the Miwa locus, which still remains a challenge in the Hermitian and complex rectangular cases. If resolved, it could extend the Hermitian superintegrability to non-Gaussian cases and to Dijkgraaf-Vafa phases \cite{DV}.

\section{Conclusion}

In this paper, we extended the study of superintegrability to unitary matrix models.
We showed that it works exactly in the same way as for the Hermitian and rectangular complex
models, still some new aspects of the story get revealed by this generalization.

An advantage of unitary models is that the one is in no way restricted to the Gaussian
measure, and $\rho(z)$ from (\ref{mainH}), while ignored in (\ref{main}),
can  easily be made non-trivial by choosing appropriate values of $\bar p_k$.
This can be important for further generalizations, say to  torus knot model
\cite{BEM,AMMM},
where $\rho(z) = \exp(\log^2 z)$, and also the Vandermonde factor in the measure is essentially modified,
which needs more work to handle.

The partition function of unitary models can be always expanded into Schur polynomials,
without a restriction to Gaussian measures needed so far in the Hermitian case.
However, this decomposition includes an explicit dependence on $N$ through the
restriction $l_R\leq N$ in the sum over partitions $R$.
This restriction can be lifted if the symmetry between $U$ and $U^\dagger$
is broken, and $\bar p_k$ are restricted to a peculiar Miwa locus $\bar p_k = \tr \bar X^k$
with $N\times N$ matrix $X$,
thus the strong $N$-dependence persists.
This is exactly in parallel with Hermitian case, still the $N$ dependence
now has two different equally efficient descriptions.

Superintegrability can be easily extended from the simplest unitary model
to many other more sophisticated examples, see ss.5--8 above.

Thus the study of unitary models confirms universality of superintegrability,
i.e. its applicability to more and more relevant models.
At the same time, it raises new questions and can help to better understand
this mysteriously general property.

\section*{Acknowledgements}

This work was supported by the Russian Science Foundation (Grant No.20-12-00195).

\section*{Appendix}

Throughout the paper, we use the skew Schur functions $S_{R/Q}\{p\}$ \cite{Mac} defined by
\be
S_R\{p+p'\}=\sum_Q S_{R/Q}\{p'\}S_Q\{p\}
\ee
and the Q Schur functions \cite{Schur, Mac}, which are defined as the Hall-Littlewood polynomials at the value of parameter $t=-1$:
\be
Q_R=
\left\{\begin{array}{cl}
2^{l_{_R}/2}\cdot\hbox{HL}_R(t=-1)&\hbox{for}\ R\in\hbox{SP}\\
&\\
0&\hbox{otherwise}
\end{array}
\right.
\ee
where $SP$ means strict partitions, i.e. those with all lines of distinct lengths. We fix their normalization as in \cite{MMkon} by the Cauchy identity
\be
\sum_RQ_R\{p\}Q_R\{p'\}=\exp\left(\sum_k{p_kp_k'\over k+1/2}\right)
\ee

At last, we use the shifted Schur functions $S_R^*\{p\}$, which can be unambiguously expressed through the shifted power sums \cite{MMNspin}
\be
p^*_k:=\sum_i \left[(x_i-i)^k-(-i)^k\right]
\ee
if one requires
\be
S^*_\mu\{p^*_k\}=S_\mu\{p^*\}+\sum_{\lambda:\ |\lambda|<|\mu|} c_{\mu\lambda}S_\lambda\{p^*_k\}\nn\\
\nn\\
S^*_\mu(R_i)=0\ \ \ \ \ \ \ \ \hbox{if }\mu\notin R
\ee
Their other definitions and properties can be found in \cite{Ok}.

\end{document}